\documentclass[amsmath,aps,pra,twocolumn]{revtex4-1}
\usepackage{graphicx}
\usepackage{times}
\usepackage{slashed}
\usepackage{bm}

\usepackage{color}

\begin{document}
\title{Thermal radiative corrections to hyperfine structure of light hydrogen-like systems}
\author{T. Zalialiutdinov$^{1,2}$, D. Glazov$^1$ and D. Solovyev$^1$}

\affiliation{ 
$^1$ Department of Physics, St. Petersburg State University, Petrodvorets, Oulianovskaya 1, 198504, St. Petersburg, Russia\\
$^2$ Petersburg Nuclear Physics Institute named by B.P. Konstantinov of National Research Centre "Kurchatov Institut", St. Petersburg, Gatchina 188300, Russia\\
}

\begin{abstract}
In this work, we consider the thermal correction to the hyperfine interaction in hydrogen, deuterium, and the $^3$He$^+$ ion. This correction is effectively described by one-loop Feynman graphs in the framework of the quantum electrodynamics theory for bound states at a finite temperature. A simple analysis shows the importance of the obtained results for future prospects for measuring hyperfine splitting. In addition, the application for testing the time variation of fundamental constants is briefly discussed.
\end{abstract}

\maketitle
\section{Introduction}

The formation and development of quantum theory for bound states is closely connected with precision spectroscopic experiments performed on the hydrogen atom and a number of hydrogen-like ions. Having their own methodological features, such systems are most accurately theoretically described within the framework of quantum electrodynamics (QED).
In the last decades, experimental measurements and theoretical calculations of the hyperfine structure (HFS) \cite{seelig:98:prl, beiersdorfer:01:pra, shabaev:94:jpb, shabaev:97:pra, blundell:97:pra, sunnergren:98:pra, Rothery, karshenboim:01, volotka:08:pra, volotka:12:prl, puchalski:13:prl, diermaier:17:n, puchalski:20:prl, puchalski:21:prr} and g-factors \cite{haefner:00:prl, sturm:13:pra, wagner:13:prl, nature2014, koehler:16:nc, harman:18:jpcs, arapoglou:19:prl, glazov:19:prl, sailer:22:n} in simple atomic systems have attracted particular interest. Significant progress in evaluation of QED radiative and nuclear-structure corrections (see, e.g., \cite{volotka:13:ap, Glazov2014, shabaev:2015:031205, indelicato:19:jpb} and references therein), as well as the improvement of experimental techniques, make it possible to identify the fundamental parameters of the theory (the electron to proton mass ratio $m_{e}/m_{p}$ \cite{massjent, nature2014}, the Rydberg constant \cite{science2020}, nuclear parameters \cite{koehler:16:nc, sailer:22:n}, etc.), to set constraints on the variation of the fine structure constant \cite{Kaplinghat, safronova:19:ap} and to look for manifestations of new physics \cite{debierre:20:plb}.

To date, the hyperfine splitting energies of atomic states are measured with extremely high accuracy and also serve as a tool for building atomic clocks, metrological frequency standards, tests of perturbative quantum chromodynamics (QCD) \cite{earlyquark} and others. In particular, the relative error in measuring the energy of the hyperfine splitting of the ground state in hydrogen is about $10^{-13}$ \cite{Hellwig}.

In view of the achieved experimental and theoretical accuracy, studies of such quantities as, e.g., hyperfine splitting, should necessarily take into account the influence of thermal radiation, the blackbody radiation (BBR) field, represented by the Planck equilibrium distribution function \cite{riehle2006frequency}. The most revealing manifestations of the effects induced by BBR are the quadratic ac-Stark shift of atomic energy levels and dynamic corrections to it, as well as line broadening due to electron transitions between atomic states stimulated by BBR \cite{Farley, GC}. These effects are extremely important for the spectroscopy of Rydberg atoms, the construction of atomic clocks, and the determination of frequency standards \cite{Hall, ItanoBBRZ, SKC}. 

The study of the effect of equilibrium thermal radiation on the characteristics of atomic systems is usually limited to the quantum mechanical (QM) description, in which the root-mean-squared field induced by BBR is considered as a perturbation \cite{Farley}. However, the QED theory makes it possible to reveal the effects that go beyond the QM approach. In our recent works, the quantum electrodynamics theory for bound states at finite temperature (BS-TQED) was developed to calculate thermal effects in atomic systems \cite{SLP-QED,S-2020,SZA-2020,SZA_2021}. Within the framework of this theory, based on the $S$-matrix formalism, various corrections to transition probabilities and ionization/recombination cross sections were also considered \cite{SLP-QED, jphysb2017, ZSL-1ph, ZAS-2l, SZATL,gfactor2022}. In this work, using previously developed methods for calculating finite temperature effects \cite{S-2020, gfactor2022}, we study thermal one-loop radiative corrections to the hyperfine splitting interval of $s$-states in hydrogen, deuterium and singly ionized helium-3 isotope.

The paper is organized as follows. In section~\ref{s1} within the framework of rigorous quantum electrodynamics at finite temperature and the adiabatic $S$-matrix formalism, equations for radiative corrections of thermal self-energy to hyperfine splitting in an one-electron ion are analytically derived. Then the results of the numerical evaluation are discussed in section~\ref{s2}. The relativistic units $ \hbar=c=m_{e}=1 $ ($ \hbar $ is the Planck constant, $ c $ is the speed of light, $ m_{e} $ is the electron mass, it is written explicitly in some places for clarity) and charge units $\alpha=e^2$ ($\alpha$ is the fine structure constant) are used throughout the paper. The product of the Boltzmann constant $ k_{B} $ and the temperature $ T $ is written in relativistic units.

\section{Basic equations for thermal loop correction to HFS splitting}
\label{s1}

The magnetic dipole moment of the nucleus is
\begin{eqnarray}
\label{1}
\bm{\mu}=g_{I}\mu_{N}\bm{I}
\end{eqnarray}
where $g_{I}$ denotes the nuclear g-factor, 	$ \mu_{N}=|e|/(2m_{p})=\mu_{B}(m_{e}/m_{p}) $ is the
nuclear magneton, $\mu_{B}=|e|/(2m_{e})$ is the Bohr magneton, and $m_e$ and $m_{p}$ are the electron and proton masses, respectively. The vector potential generated by the nuclear dipole moment given by Eq. (\ref{1}) is
\begin{eqnarray}
\label{2}
\bm{A}=\frac{\bm{\mu}\times \bm{r}}{r^3}
.
\end{eqnarray}
Then the interaction of the bound electron with the dipole nuclear magnetic field is given by the Fermi-Breit operator
\begin{eqnarray}
\label{3}
\hat{V}_{\mathrm{hfs}} = -e \bm{\alpha}\bm{A}=|e|\frac{\bm{\alpha}\cdot(\bm{\mu}\times \bm{r})}{r^3} 
,
\end{eqnarray}
where $\bm{\alpha}$ is the Dirac alpha matrix. 

The corresponding expectation value of Eq. (\ref{3}) on the Dirac point-nucleus wave functions of a hydrogen-like ion looks as follows
\begin{eqnarray}
\label{4}
E_{\mathrm{hfs}}=\alpha(\alpha Z)^3\frac{g_{I}}{2}\frac{m_{e}^2}{m_{N}}\frac{\kappa}{|\kappa|}\frac{1}{n^3(2\kappa + 1)(\kappa^2 -1/4)}
\\\nonumber
\times A( Z \alpha)[F(F+1)-I(I+1)-j(j+1)]
,
\end{eqnarray}
where 
$ A( Z \alpha)$ is the relativistic factor
\begin{eqnarray}
\label{5}
A( Z \alpha) = n^3|\kappa|(2\kappa +1)\frac{2\kappa(2\gamma + n_{r})-N}{N^4\gamma (4\gamma^2 - 1)}
.
\end{eqnarray}
Here,~$N=\sqrt{n_{r}^2+2n_{r}\gamma + \kappa^2}$,~$n_{r}=n-|\kappa|$,~$\gamma=\sqrt{\kappa^2-(\alpha Z)^2}$,~ $n$ is the principal quantum number of the electron,~$\kappa$ is its Dirac angular quantum number,~$j=|\kappa|-1/2$, $Z$ is the nuclear charge, $F$ is the total angular momentum of atom and $I$ is the nuclear spin.

In the nonrelativistic limit, the hyperfine Hamiltonian is given by the sum of two contributions. The first one is proportional to $\bm{s}\cdot \bm{B}$, where $\bm{s}$ is the electron spin and $\bm{B} $ is the magnetic field corresponding to the vector potential Eq.~(\ref{2}): 
\begin{eqnarray}
\label{6}
\bm{B}=\nabla\times\bm{A}=\frac{8\pi}{3}\bm{\mu}\delta^{(3)}(\bm{r}) + \frac{1}{r^3}
\left(3\left(\bm{\mu}\cdot\bm{n}\right)\bm{n}-\bm{\mu}\right)
.
\end{eqnarray}
 The second contribution corresponds to the interaction of the nuclear moment with the magnetic field of the moving electron, which in turn is proportional to the orbital angular momentum $\bm{l}$. Then in the nonrelativistic limit we have
\begin{eqnarray}
\label{7}
\hat{V}_{\mathrm{hfs}}^{\mathrm{nr}} = -2\mu_{B}
\left[-
\frac{8\pi}{3}
(\bm{s}\cdot\bm{\mu})\delta^{(3)}(\bm{r})
\right.
\\\nonumber
-
\frac{\bm{l}\cdot \bm{\mu}}{r^3}
\left.
-\frac{1}{r^3}
\left(3\left(\bm{s}\cdot\bm{n}\right)\bm{n}-\bm{s}\right)
\cdot
\bm{\mu}
\right]
.
\end{eqnarray}
The evaluation of matrix elements of Eq. (\ref{7}) with Schr\"odinger point-nucleus wave functions in the $ lsj(I)FM_{F} $ coupling scheme is presented in Appendix A. For the hyperfine splitting of the ground state in hydrogen, using the expression (\ref{7}), we find the known wavelength value 21 cm, or 1420 MHz for the frequency. 

\begin{figure}[hbtp]
\centering
\includegraphics[scale=2]{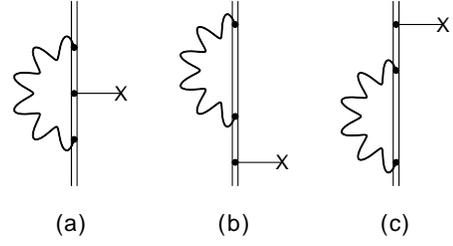}
\caption{Feynman diagrams describing the TQED self-energy corrections to the hyperfine interaction vertex. The tiny line with the cross indicates interaction with vector potential generated by the nuclear dipole moment. The double solid line denotes the bound electron in the Furry picture. The bold wavy line represents the thermal photon propagator.}
\label{fig1}
\end{figure}

The thermal self-energy (TSE) correction in the presence of a binding Coulomb field and an additional perturbing potential defined by Eq. (\ref{3}) is graphically represented by the Feynman diagrams in Fig. \ref{fig1}. The total energy shift of the third order in $ e $ can be written as a sum of vertex (Fig. \ref{fig1} (a)) and wave-function (Fig. \ref{fig1} (b), (c)) contributions, $\Delta E_{a}^{\mathrm{total}}=\Delta E_{a}^{\mathrm{ver}}+\Delta E_{a}^{\mathrm{wf}}$, see e.g. \cite{gfactor2022} for details, as follows
\begin{eqnarray}
\label{11}
\Delta E_{a}^{\mathrm{ver}}
=-\frac{e^3 }{\pi}\mathrm{Re}
\sum\limits_{\pm}
\sum\limits_{n,m}
\int\limits_{0}^{\infty}
d\omega n_{\beta}(\omega)\qquad
\\\nonumber
\times
\frac{\langle am| \frac{\alpha^{\mu}_{1}\alpha_{2\mu}}{r_{12}} \sin(\omega r_{12}) |na\rangle
\langle n| \bm{\alpha}\bm{A}|m \rangle 
}{(E_{a}\pm\omega-E_{n}(1-\mathrm{i}0))(E_{a}\pm\omega-E_{m}(1-\mathrm{i}0))}
\,,
\end{eqnarray}
\begin{eqnarray}
\label{12}
\Delta E_{a}^{\mathrm{wf}}
=-\frac{2e^3 }{\pi}\mathrm{Re}
\sum\limits_{\pm}\int\limits_{0}^{\infty}
d\omega n_{\beta}(\omega)\qquad
\\
\nonumber
\times
\left[
\sum\limits_{\substack{n,m\\m\neq a}}
\frac{\langle an| \frac{\alpha^{\mu}_{1}\alpha_{2\mu}}{r_{12}} \sin(\omega r_{12}) |nm\rangle
\langle m| \bm{\alpha}\bm{A}|a \rangle 
}{(E_{a}\pm\omega-E_{n}(1-\mathrm{i}0))(E_{a}-E_{m})}
\right.
\\\nonumber
\left.
-\frac{1}{2}
\sum\limits_{\substack{n}}
\frac{\langle an| \frac{\alpha^{\mu}_{1}\alpha_{2\mu}}{r_{12}} \sin(\omega r_{12}) |na\rangle
\langle a| \bm{\alpha}\textbf{A}|a \rangle 
}{(E_{a}\pm\omega-E_{n}(1-\mathrm{i}0))^2}
\right]
\,.
\end{eqnarray}

Here $\sum\limits_{\pm}$ denotes the sum of two contributions with $+$ and $-$ in energy denominators, vector potential $\bm{A}$ is given by Eq. (\ref{2}), $ n_{\beta}(\omega)=(\exp(\omega/(k_{B}T)-1)^{-1} $, $k_{B}$ is the Boltzmann constant, $T$ is the temperature (in Kelvin). In contrast to the ordinary 'zero-vacuum' corrections, the expressions (\ref{11}) and (\ref{12}) converge in the ultraviolet limit due to the natural cut-off provided by the Planckian distribution function $ n_{\beta}(\omega) $. The second term in Eq. (\ref{12}) corresponds to the "reference-state" contribution, $ m=a $.

Then substituting Eq. (\ref{2}) into Eqs. (\ref{11}) and (\ref{12}) and taking the nonrelativistic limit (see details in \cite{gfactor2022}), we find
\begin{align}
\label{13}
\Delta E_{a}^{\mathrm{ver}}
=-\frac{2 e^2 }{3\pi}\mathrm{Re}
\sum\limits_{\pm}
\sum\limits_{n,m}
\int\limits_{0}^{\infty}
d\omega \omega^3 n_{\beta}(\omega)\qquad
\\\nonumber
\times
\frac{\langle a| \textbf{r} |n\rangle \langle n|\hat{V}_{\mathrm{hfs}}^{\mathrm{nr}}|m \rangle \langle m|\textbf{r}|a\rangle
}{(E_{a}\pm\omega-E_{n}(1-\mathrm{i}0))(E_{a}\pm\omega-E_{m}(1-\mathrm{i}0))}
,
\end{align}
\begin{align}
\label{14}
\Delta E_{a}^{\mathrm{wf}}
=-\frac{4e^2  }{3\pi}
\mathrm{Re}\sum\limits_{\pm}
\int\limits_{0}^{\infty}
d\omega n_{\beta}(\omega)
\omega^3\qquad
\\\nonumber
\times
\left[
\sum\limits_{\substack{n,m\\m\neq a}}
\frac{\langle a|\textbf{r}|n\rangle \langle n|\textbf{r}|m\rangle \langle m| \hat{V}_{\mathrm{hfs}}^{\mathrm{nr}}|a \rangle 
}{(E_{a}\pm\omega-E_{n}(1-\mathrm{i}0))(E_{a}-E_{m})}
\right.
\\\nonumber
-
\left.
\frac{1}{2}
\sum\limits_{n}
\frac{\langle a|\textbf{r}|n\rangle \langle n|\textbf{r}|a\rangle \langle a|\hat{V}_{\mathrm{hfs}}^{\mathrm{nr}}|a \rangle 
}{(E_{a}\pm\omega-E_{n}(1-\mathrm{i}0))^2}
\right]
,
\end{align}
where $\hat{V}_{\mathrm{hfs}}^{\mathrm{nr}}$ is given by Eq. (\ref{7}) and the summation over the $n$ spectrum is carried out over the discrete and continuum solutions of the Schr\"odinger equation for an electron in the Coulomb field of the nucleus.

The corresponding total shift for the transition energy between hyperfine components $F_{a}'$ and $F_{a}$ of the same state $n_{a}l_{a}j_{a}$ is
\begin{eqnarray}
\label{15}
\Delta\nu^{\mathrm{hfs}}=\Delta E_{n_al_as_aj_aF'_a}^{\mathrm{total}} - \Delta E_{n_al_as_aj_aF_a}^{\mathrm{total}} 
.
\end{eqnarray}
For hydrogen-like ions with a nuclear charge $Z$, a parametric estimates for Eqs. (\ref{13}) and (\ref{14}) can be found taking into account that in relativistic units $ r\sim (m_{e}\alpha Z)^{-1} $, $ E_{a}\sim m_{e}(\alpha Z)^2 $ and
$\int\limits_{0}^{\infty}d\omega\,\omega^{k}n_{\beta}(\omega)\sim(k_{B}^{\mathrm{\;r.u.}}T)^{k+1}$. Then $\Delta\nu^{\mathrm{hfs}}$ given by Eq. (\ref{15}) is parametrized as follows
\begin{eqnarray}
\label{est1}
\Delta\nu^{\mathrm{hfs}}\sim \frac{(k_{B}T)^4_{\mathrm{r.u.}}}{\alpha m_{p} m_{e}^2 Z^{3}}
\,
,
\end{eqnarray}
where $m_{p}$ is the proton mass.

In particular, the estimate Eq. (\ref{est1}) is valid for the state $ a=1s $, when the summation runs over the $ np $ states and the energy difference in the denominators of Eqs. (\ref{13}) and (\ref{14}) is always of the order $ m_{e}(\alpha Z)^2 $. However, for $ n_{a}l_{a} $ states with $ n_{a} \geq 2 $ (for example, $a=2s$), the dominant contribution in the sum over $ n $ corresponds to $ E_{n}=E_{n_{a}} $. In this case, the following parametrization is valid
\begin{eqnarray}
\label{est2}
\Delta\nu^{\mathrm{hfs}}\sim  \frac{Z\alpha^3(k_{B}T)^2_{\mathrm{r.u.}}}{m_{p}}
\,.
\end{eqnarray}

\section{Results and discussion}
\label{s2}

Numerical results of $\Delta\nu^{\mathrm{hfs}}$ calculation for $ns_{1/2}$ ($n=1,\,2,\,3,\,4$) states in hydrogen, deuterium and $^3$He$^+$ ion are collected in Tables~\ref{tab1}, \ref{tab2}, and \ref{tab3HE}, respectively, at different temperatures. Numerical evaluation of radial integrals and summation over the entire spectrum in Eqs. (\ref{13}) and (\ref{14}) was performed using the B-spline method for the solution of Schr\"odinger equation. 
\begin{table}[ht]
\caption{Thermal self-energy correction to the hyperfine splitting  $\Delta\nu^{\mathrm{hfs}}_{a}$ of $ns$, $np$ and $nd$ states in hydrogen $e^-p^+$ ($I=1/2$, $g_{I}=5.5856946893(16)$) at different temperatures (in kelvin). All values are in hertz.}
\label{tab1}
\begin{tabular}{c c c c}
\hline
\mbox{Energy shift} & T=300 & T=1000 & T=3000 \\
\hline
$ 1s_{1/2}^{F=1-F=0} $ & $2.00\times 10^{-8}$   & $2.48\times 10^{-6}$  & $2.07\times 10^{-4}$\\
$ 2s_{1/2}^{F=1-F=0}  $ & $1.17\times 10^{-3}$  & $1.31\times 10^{-2}$  & $1.23\times 10^{-1}$\\
$ 2p_{1/2}^{F=1-F=0}  $ & $2.61\times 10^{-4}$  & $2.93\times 10^{-3}$  & $2.88\times 10^{-2}$\\
$ 2p_{3/2}^{F=2-F=1}  $ & $-2.09\times 10^{-4}$ & $ -2.32\times 10^{-3}$& $-1.96\times 10^{-2}$\\
$ 3s_{1/2}^{F=1-F=0}  $ & $2.10\times 10^{-3}$  & $2.39\times 10^{-2}$  & $1.85\times 10^{-1}$\\
$ 3p_{1/2}^{F=1-F=0}  $ & $6.97\times 10^{-4}$  & $7.90\times 10^{-3}$  & $6.15\times 10^{-1}$\\
$ 3p_{3/2}^{F=2-F=1}  $ & $-3.06\times 10^{-5}$ & $3.52\times 10^{-4}$  & $-7.71\times 10^{-5}$\\
$ 3d_{3/2}^{F=2-F=1}  $ & $-8.72\times 10^{-5}$ & $-8.75\times 10^{-4}$ & $-1.42\times 10^{-2}$\\
\hline
\end{tabular}
\end{table}
\begin{table}[ht]
\caption{Thermal self-energy correction to the hyperfine splitting $\Delta\nu^{\mathrm{hfs}}$ of $ns$, $np$ and $nd$ states in deuterium $e^-d^+$ ($I=1$, $g_{I}=0.8574382338(22)$) at different temperatures (in kelvin). All values are in hertz.}
\label{tab2}
\begin{tabular}{c c c c}
\hline
\mbox{Energy shift} & T=300 & T=1000 & T=3000 \\
\hline
$ 1s_{1/2}^{F=3/2-F=1/2}  $ & $ 4.61\times 10^{-9} $ & $ 5.70\times 10^{-7} $& $ 4.76\times 10^{-5} $\\
$ 2s_{1/2}^{F=3/2-F=1/2}  $ & $ 2.71\times 10^{-4} $ & $ 3.02\times 10^{-3} $& $ 2.82\times 10^{-2}$\\
$ 2p_{1/2}^{F=3/2-F=1/2}  $ & $ 6.01\times 10^{-5}$ & $ 6.74\times 10^{-4} $& $ 6.65\times 10^{-3}$\\
$ 2p_{3/2}^{F=3/2-F=1/2}  $ & $ -2.40\times 10^{-5}$ & $  -2.67\times 10^{-4} $& $ -2.36\times 10^{-3}$\\
$ 2p_{3/2}^{F=5/2-F=3/2}  $ & $  -4.01\times 10^{-5}$ & $  -4.44\times 10^{-4} $& $ -3.60\times 10^{-3}$\\
$ 3s_{1/2}^{F=3/2-F=1/2}  $ & $  4.81\times 10^{-4}$ & $ 5.56\times 10^{-3} $& $ 4.24\times 10^{-2} $\\
$ 3p_{1/2}^{F=3/2-F=1/2}  $ & $ 1.61\times 10^{-4}$ & $ 5.76\times 10^{-3} $& $ 1.41\times 10^{-4}$\\
$ 3p_{3/2}^{F=3/2-F=1/2}  $ & $ -3.41\times 10^{-5}$ & $  -3.66\times 10^{-4} $& $ -4.48\times 10^{-3}$\\
$ 3p_{3/2}^{F=5/2-F=3/2}  $ & $ -5.69\times 10^{-5}$ & $ -6.05\times 10^{-4}  $& $ -7.51\times 10^{-3}$\\
$ 3d_{3/2}^{F=3/2-F=1/2}  $ & $ -8.45\times 10^{-6}$ & $ -1.01\times 10^{-4}$ & $ -9.58\times 10^{-4}$\\
$ 3d_{3/2}^{F=5/2-F=3/2}  $ & $-1.41\times 10^{-5}$ & $-1.68\times 10^{-4}$ & $-1.69\times 10^{-3}$\\
\hline
\end{tabular}
\end{table}
\begin{table}[ht]
\caption{Thermal self-energy correction to the hyperfine splitting $\Delta\nu^{\mathrm{hfs}}$ of $ns$, $np$ and $nd$ states in $^3$He$^+$ ($I=1/2$, $g_{I}=-4.255250615(50)$) at different temperatures (in kelvin). All values are in hertz.}
\label{tab3HE}
\begin{tabular}{c c c c}
\hline
\mbox{Energy shift} & T=300 & T=1000 & T=3000 \\
\hline
$ 1s_{1/2}^{F=1-F=0}  $ & $  -1.90\times 10^{-9} $ & $ -2.35\times 10^{-7} $& $ -1.91\times 10^{-5} $\\
$ 2s_{1/2}^{F=1-F=0}  $ & $  -1.79\times 10^{-3} $ & $ -1.99\times 10^{-2}  $& $ -1.79\times 10^{-1} $\\
$ 2p_{1/2}^{F=1-F=0}  $ & $  -3.98\times 10^{-4} $ & $ -4.42\times 10^{-3}  $& $ -4.00\times 10^{-2} $\\
$ 2p_{3/2}^{F=2-F=1}  $ & $  3.18\times 10^{-4} $ & $ 2.33\times 10^{-3}  $& $ 3.19\times 10^{-2} $\\
$ 3s_{1/2}^{F=1-F=0}  $ & $  -3.18\times 10^{-3} $ & $ -3.54\times 10^{-2}  $& $ -3.23\times 10^{-1} $\\
$ 3p_{1/2}^{F=1-F=0}  $ & $  -1.06\times 10^{-3} $ & $ -1.18\times 10^{-2}  $& $ -1.09\times 10^{-1} $\\
$ 3p_{3/2}^{F=2-F=1}  $ & $  1.30\times 10^{-4} $ & $ 1.45\times 10^{-3}  $& $ 1.36\times 10^{-2} $\\
$ 3d_{3/2}^{F=2-F=1}  $ & $  1.33\times 10^{-4} $ & $ 1.48\times 10^{-3}  $& $ 1.31\times 10^{-2} $\\
\hline
\end{tabular}
\end{table}

Due to the uncertainty in determining the size of nuclei, the possibility of carrying out QED tests in atomic systems based on the analysis of a specific energy difference has been proposed \cite{Zwanziger, shabaev:01:prl}. A simple analogue of this difference for light systems, which is weakly sensitive to the contributions arising from the nuclear structure, is
\begin{eqnarray}
\label{diff}
D_{21}=8\Delta\nu^{\mathrm{hfs}}_{2s}-\Delta\nu^{\mathrm{hfs}}_{1s}
\,.
\end{eqnarray}

Over the past few years, the accuracy of calculating corrections to the energy of hyperfine splitting and the value of $D_{21}$ has increased significantly. A number of corrections have been calculated, including those for the electroweak interaction, the hadronic vacuum polarization, and the structure of the nucleus \cite{Alcorta1994, Asaka, Faustov1999, Beg, Kalinowski}. 
At present, the inaccuracy of theoretical calculations is much lower than the experimental error in measuring the parameter $D_{21}$, which is mainly determined by the error in measuring the hyperfine splitting of the $2s$ metastable state. For hydrogen and deuterium atoms, the experimental error in measuring the $2s$ hyperfine interval is about ten hertz \cite{Kusch, Fendel}, while for the ground state it reaches several millihertz \cite{Hellwig, Ramsey, Mathur}. 

According to the results listed in Tables~\ref{tab1}, \ref{tab2} and \ref{tab3HE}, the correction Eq. (\ref{15}) for the ground state can be excluded from consideration for the $D_{21}$ difference in hydrogen, deuterium and $^3$He$^+$ ion. However, a decrease in the experimental error in determining the hyperfine splitting of the $2s$ state to the accuracy level of the ground state splitting will be sensitive to this correction.

Another application of the correction considered in this paper concerns the search for time variation of fundamental physical constants, see, for example, \cite{Webb,Kaplinghat}. Verification of the fine structure constant variation can be performed by comparing in detail the spectral data of quasars and laboratory results for the hyperfine splitting of the ground state in the hydrogen atom. In this case, the higher temperatures can play a role. According to the results listed in Table~\ref{tab1}, the $\Delta\nu^{\mathrm{hfs}}_{a}$ reaches the accuracy level of laboratory experiments only at sufficiently high $3000$ K and become larger at higher temperatures. Thus, corrections (\ref{13}) and (\ref{14}) impose additional constrains on this type of research. Simple analysis of calculated shifts in hydrogen, deuterium and $^3$He$^+$ presented in Tables \ref{tab1}, \ref{tab2} and \ref{tab3HE} shows that hyperfine splitting interval is weakly sensitive to the considered thermal correction.

Despite the obtained values are so small, it is important to note that the thermal one-loop correction to the hyperfine structure considered in this paper is of the order of Zeeman shifts induced by blackbody radiation (BBRZ) \cite{Han2019, ItanoBBRZ}. For alkali metals and alkali-like ions the relative value of the BBRZ shift for the ground state reaches the same order of magnitude as for hydrogen (see Table 2 in \cite{Han2019}). For many-electron atoms and ions with one valence electron, rough estimates can be given with the value of $Z_{eff}$ little different from unity \cite{Clementi}. Thus, using estimates (\ref{13}), (\ref{14}) when $Z$ is replaced by $Z_{eff}$, we can expect a contribution of the same order for alkali metals. This can be especially important when elaborating frequency standards operating on hyperfine transitions of the outermost $s$-electron. We leave detailed calculations of thermal one-loop corrections in alkali metals for future work.

\section{Acknowledgements}
The calculations for hydrogen and deuterium were supported by the Russian Scientific Foundation under grant No. 22-12-00043.
The calculations for helium were supported by the Russian Scientific Foundation under grant No. 22-12-00258.

\appendix
\renewcommand{\theequation}{A\arabic{equation}}
\setcounter{equation}{0}

\section*{Appendix A: Evaluation of matrix elements}
Taking into account Eq. (\ref{1}) and performing angular algebra with the use of Eckart-Wigner theorem the first term in Eq. (\ref{7}) can be evaluated with the help of following equality
\begin{eqnarray*}
\langle n'l's'j'(I) F'M_{F'}| (\bm{s}\cdot\bm{I}) \delta^3(\bm{r})|nlsj(I)FM_{F}\rangle
\\\nonumber
=
\delta_{F'F}\delta_{M_{F'}M_{F}}\delta_{l'l}\delta_{s's}(-1)^{F+j+I+l'+s+j'+1}
\\\nonumber
\times
\sqrt{\frac{3}{2}}
\begin{Bmatrix}
j' & j & 1\\
I  & I & F
\end{Bmatrix}
\begin{Bmatrix}
s' & j' & l\\
j  & s  & 1
\end{Bmatrix}
\\\nonumber
\times
\left[
(2j'+1)
(2j+1)I(I+1)(2I+1)
\right]^{1/2}
\frac{1}{4\pi} R_{n'l'}(0)R_{nl}(0)
.
\end{eqnarray*}
The radial nonrelativistic wave-functions $R_{nl}$ taken at zero are nonvanishing only for $s$-states. Then for discrete and continuum solutions of Schr\"odinger equation we have
\begin{eqnarray}
R_{n0}(0)=\frac{2}{n^{3/2}},
\\\nonumber
R_{x0}(0)=\frac{2\sqrt{x}}{\sqrt{1-e^{-2\pi/x}}}
,
\end{eqnarray}
respectively. 

In a similar manner, the second term in Eq. (\ref{7}) can be evaluated with the use of equality
\begin{eqnarray*}
\langle n'l's'j'(I) F'M_{F'}| \frac{\bm{l}\cdot \bm{I}}{r^3} |nlsj(I)FM_{F}\rangle
\\\nonumber
=\delta_{F'F}\delta_{M_{F'}M_{F}}\delta_{l'l}\delta_{s's}
(-1)^{2j+l'+s'+I+F+1}
\\\nonumber
\times
\left[
(2j'+1)
(2j+1)I(I+1)(2I+1)
\right]^{1/2}
\\\nonumber
\times
\begin{Bmatrix}
F  & I & j'\\
1  & j & I
\end{Bmatrix}
\begin{Bmatrix}
l' & j' & s'\\
j  & l  & 1
\end{Bmatrix}
\sqrt{l(l+1)(2l+1)}
\\\nonumber
\times
\int\limits_{0}^{\infty}dr r^2R_{n'l'}(r)\left(\frac{1}{r^3}\right)R_{nl}(r)
.
\end{eqnarray*}
The third term in Eq. (\ref{7}) can be easily evaluated with the use of following relation
\begin{eqnarray}
\bm{K}\cdot\bm{I}
\equiv 
r^{-3}\left(3\left(\bm{s} \cdot\bm{n}\right)\bm{n} - \bm{s}\right)
\cdot
\bm{I} 
\\\nonumber
 = r^{-3}\sqrt{10}\sum\limits_{q}(-1)^q [ \mathrm{C}^{2}\times \mathrm{S}^{1}]_{1q}I_{1-q}
 .
\end{eqnarray}
Then the reduction of the corresponding matrix elements yields
\begin{eqnarray}
\langle n'l's'j'(I)F'M_{F'}| \bm{K}\cdot\bm{I} |nlsj(I)FM_{F}\rangle\qquad
\\\nonumber
=\delta_{F'F}\delta_{M_{F'}M_{F}}\delta_{s's}
\sqrt{45}(-1)^{j+I+F+l'}
\begin{Bmatrix}
F & I & j'\\
1 & j & I
\end{Bmatrix}
\\\nonumber
\times
[I(I+1)(2I+1)(2l'+1)(2l+1)
(2j'+1)(2j+1)]^{1/2}
\\\nonumber
\times
\begin{Bmatrix}
l' & l & 2\\
s' & s & 1\\
j' & j & 1
\end{Bmatrix}
\begin{pmatrix}
l' & 2 & l\\
0  & 0 & 0
\end{pmatrix}
\int\limits_{0}^{\infty}dr r^2R_{n'l'}(r)\left(\frac{1}{r^3}\right)R_{nl}(r)
.
\end{eqnarray}

\bibliography{mybibfile}

\end{document}